\documentstyle[preprint,psfig,aps,floats,tighten]{revtex} 

\begin{document}

\preprint{FERMILAB-Pub-97/109-E}
\title{Measurement of the Top Quark Pair Production Cross Section
in $p\overline p$ Collisions}

\author{
\centerline{The D\O\ Collaboration\thanks{Authors listed on the following page.
            \hfill\break 
            Submitted to Physical Review Letters.}}
}

\address{
\centerline{Fermi National Accelerator Laboratory, Batavia, Illinois 60510}
}

\date{April 22, 1997}

\maketitle

\begin{abstract}
We present a measurement of the $t \overline t$ production cross section
in $p\bar p$ collisions at $\sqrt{s} = 1.8$~TeV by the D\O\ experiment
at the Fermilab  Tevatron. The measurement is based on data from an
integrated luminosity of approximately 125~pb$^{-1}$ accumulated during
the 1992--1996 collider run. We observe 39 $t \overline t$ candidate
events in the dilepton and lepton+jets decay channels with an expected
background of \hbox{$13.7\pm2.2$} events.  For a top quark mass of
$173.3$~GeV/c$^2$, we measure the $t \overline t$ production cross
section to be $5.5 \pm 1.8$~pb.
\end{abstract}
\pacs{PACS numbers 14.65.Ha, 13.85.Qk, 13.85.Ni}

\vskip 1cm

\begin{center}
S.~Abachi,$^{14}$                                                             
B.~Abbott,$^{28}$                                                             
M.~Abolins,$^{25}$                                                            
B.S.~Acharya,$^{43}$                                                          
I.~Adam,$^{12}$                                                               
D.L.~Adams,$^{37}$                                                            
M.~Adams,$^{17}$                                                              
S.~Ahn,$^{14}$                                                                
H.~Aihara,$^{22}$                                                             
G.A.~Alves,$^{10}$
% ...
\end{center}
\vskip 0.50cm
\normalsize
\centerline{(D\O\ Collaboration)}

\vfill\eject
\small
\it
\centerline{$^{1}$Universidad de los Andes, Bogot\'{a}, Colombia}             
\centerline{$^{2}$University of Arizona, Tucson, Arizona 85721}               
\centerline{$^{3}$Boston University, Boston, Massachusetts 02215}             
\centerline{$^{4}$Brookhaven National Laboratory, Upton, New York 11973}      
\centerline{$^{5}$Brown University, Providence, Rhode Island 02912}           
\centerline{$^{6}$Universidad de Buenos Aires, Buenos Aires, Argentina}       
\centerline{$^{7}$University of California, Davis, California 95616}          
\centerline{$^{8}$University of California, Irvine, California 92697}         
\centerline{$^{9}$University of California, Riverside, California 92521}      
\centerline{$^{10}$LAFEX, Centro Brasileiro de Pesquisas F{\'\i}sicas,        
                  Rio de Janeiro, Brazil}                                     
\centerline{$^{11}$CINVESTAV, Mexico City, Mexico}                            
\centerline{$^{12}$Columbia University, New York, New York 10027}             
\centerline{$^{13}$Delhi University, Delhi, India 110007}                     
\centerline{$^{14}$Fermi National Accelerator Laboratory, Batavia,            
                   Illinois 60510}                                            
\centerline{$^{15}$Florida State University, Tallahassee, Florida 32306}      
\centerline{$^{16}$University of Hawaii, Honolulu, Hawaii 96822}              
\centerline{$^{17}$University of Illinois at Chicago, Chicago,                
                   Illinois 60607}                                            
\centerline{$^{18}$Indiana University, Bloomington, Indiana 47405}            
\centerline{$^{19}$Iowa State University, Ames, Iowa 50011}                   
\centerline{$^{20}$Korea University, Seoul, Korea}                            
\centerline{$^{21}$Kyungsung University, Pusan, Korea}                        
\centerline{$^{22}$Lawrence Berkeley National Laboratory and University of    
                   California, Berkeley, California 94720}                    
\centerline{$^{23}$University of Maryland, College Park, Maryland 20742}      
\centerline{$^{24}$University of Michigan, Ann Arbor, Michigan 48109}         
\centerline{$^{25}$Michigan State University, East Lansing, Michigan 48824}   
\centerline{$^{26}$Moscow State University, Moscow, Russia}                   
\centerline{$^{27}$University of Nebraska, Lincoln, Nebraska 68588}           
\centerline{$^{28}$New York University, New York, New York 10003}             
\centerline{$^{29}$Northeastern University, Boston, Massachusetts 02115}      
\centerline{$^{30}$Northern Illinois University, DeKalb, Illinois 60115}      
\centerline{$^{31}$Northwestern University, Evanston, Illinois 60208}         
\centerline{$^{32}$University of Notre Dame, Notre Dame, Indiana 46556}       
\centerline{$^{33}$University of Oklahoma, Norman, Oklahoma 73019}            
\centerline{$^{34}$University of Panjab, Chandigarh 16-00-14, India}          
\centerline{$^{35}$Institute for High Energy Physics, 142-284 Protvino,       
                   Russia}                                                    
\centerline{$^{36}$Purdue University, West Lafayette, Indiana 47907}          
\centerline{$^{37}$Rice University, Houston, Texas 77005}                     
\centerline{$^{38}$Universidade Estadual do Rio de Janeiro, Brazil}           
\centerline{$^{39}$University of Rochester, Rochester, New York 14627}        
\centerline{$^{40}$CEA, DAPNIA/Service de Physique des Particules,            
                   CE-SACLAY, Gif-sur-Yvette, France}                         
\centerline{$^{41}$Seoul National University, Seoul, Korea}                   
\centerline{$^{42}$State University of New York, Stony Brook,                 
                   New York 11794}                                            
\centerline{$^{43}$Tata Institute of Fundamental Research,                    
                   Colaba, Mumbai 400005, India}                              
\centerline{$^{44}$University of Texas, Arlington, Texas 76019}               
\centerline{$^{45}$Texas A\&M University, College Station, Texas 77843}       

\normalsize

\vfill\eject

The discovery\cite{topobs} of the top quark in 1995 at the Fermilab
Tevatron collider ended a long search following the 1977 discovery of
the $b$ quark\cite{bdisc}  and represents another triumph of the
Standard Model (SM). In the SM, the top quark completes the third
fermion generation. A measurement of the top quark pair production cross
section is of interest as a test of QCD predictions. A deviation from
these predictions could indicate non-standard production or decay
processes.

In $p\overline p$ collisions at $\sqrt s$ = 1.8 TeV, top and anti-top
quarks are predominantly pair produced through $q\overline q$
annihilation ($\approx$90\%) or gluon fusion ($\approx$10\%). In the SM, 
due to their large mass, they decay before they hadronize;  nearly all
($\geq$99.8\%) decay to a $W$ boson and a $b$ quark. The subsequent $W$
decay  determines the major signatures of $t\overline t$ decay. In the
dilepton channel, both $W$ bosons decay  either to $e\nu$ or $\mu\nu$.
The branching fraction for this channel is rather small (4/81), but it
has the advantage of  small backgrounds. In the lepton+jets channel, one
$W$ boson decays to $e\nu$ or $\mu\nu$ and the other hadronically. The
branching fraction is 24/81. The dominant source of background for this
channel is $W$+jets production.

In this Letter we  report a measurement of the $t\overline t$ production
cross section ($\sigma_{t\overline t}$)  using the entire data sample
(125$\pm$7~pb$^{-1}$) collected during the 1992--1996 collider run. This
is more than twice the data described in  our previous
publication\cite{topobs}. Different trigger conditions cause the
integrated  luminosity to vary from channel to channel. The analysis
presented here is optimized to maximize the expected precision of the
$t\overline t$ cross section measurement.

A detailed description of the D\O\ detector, trigger, and algorithms for
reconstructing jets  and missing transverse energy 
{\hbox{$\rlap{\kern0.25em/}E_T$}} is found in Refs. \cite{D0detector}
and \cite{topprd}. The  current electron and muon identification
algorithms provide better rejection of backgrounds and increased
efficiencies than those used in Ref.\cite{topprd}.

The signature of the dilepton channel consists of two isolated high
$p_T$ leptons, two or more jets, and large 
{\hbox{$\rlap{\kern0.25em/}E_T$}}. The  selection criteria are
summarized in Table~\ref{tab:top}. Several additional cuts that remove
specific backgrounds have been omitted from the table, but are noted 
below. In Table~\ref{tab:top}, $\eta$ is the pseudorapidity, $H{_T}$ is
the scalar sum of the $E_T$ of all jets with $E_T\geq 15$ GeV, and
$H_T^e=H{_T}+E{_T}({\rm leading\ electron})$. Three $e\mu$ events, one
$ee$ event, and one $\mu\mu$ event survive the selection criteria.

\begin{table}[t]
\caption{Kinematic selection criteria for decay channels included in the cross
section measurement. An event may populate only one channel. All energies 
are in GeV.}
\label{tab:top}
\begin{center}
\begin{tabular}{lcccc}
                & dilepton          & $\ell$+jets      & $\ell$+jets/$\mu$ & $e\nu$     \\
\hline
lepton $p_T$    & $>15$             & $>20$            & $>20$             & $>20$      \\
                & $>20$ ($ee$)      &                  &                   &            \\
electron $|\eta|$ & $<2.5$          & $<2.0$           & $<2.0$            & $<1.1$     \\
muon     $|\eta|$ & $<1.7$          & $<1.7$           & $<1.7$            &  ---       \\
{\hbox{$\rlap{\kern0.25em/}E_T$}}  
       & $>20$ ($e\mu$)    & $>25$ ($e$)      & $>20$             & $>50$      \\
                & $>25$ ($ee$)      & $>20$ ($\mu$)    &                   &            \\
jet $E_T$       & $>20$             & $>15$            & $>20$             & $>30$      \\
jet $|\eta|$    & $<2.5$            & $<2.0$           & $<2.0$            & $<2.0$     \\
\# of jets      & $\geq2 $          & $\geq 4$         & $\geq 3$          & $\geq 2$   \\
$H_T^e  $       &$>120$($ee$,$e\mu$)& ---              & ---               & ---        \\
$H_T$           &$>100$ ($\mu\mu$)  & $>180$           & $>110$            & ---        \\
${\cal A}$      & ---               & {$ > 0.065$}     & {$ > 0.040$}      & ---        \\
$E_T^L$         & ---               &{ $> 60$}         & ---               & ---        \\
$\eta_W$       & ---               &{$<2.0$}          & ---               & ---        \\
tag muon        & ---               & veto             & $p_T>4$           & ---        \\
                &                   &                  & $\Delta {\cal R}_{\rm jet}<0.5$    &      \\
$M_T^{e\nu}$ 
       & ---               & ---              & ---               & $>$115     \\
\end{tabular}
\end{center}
\end{table}

The signature of  the lepton+jets channel consists of one isolated high
$p_T$ lepton, {\hbox{$\rlap{\kern0.25em/}E_T$}} due to the neutrino, and
several jets. In these events, jets are produced by the hadronization of
two $b$ quarks and the two quarks from $W$ boson decay. Thus we expect
to see four jets. However, due to gluon radiation and merging of jets,
the number of detected jets may vary. After requiring an isolated high
$p_T$ lepton,  {\hbox{$\rlap{\kern0.25em/}E_T$}}, and at least three
jets, we expect 50 events from $t\overline t$ production (assuming top
quark mass $m_t = 170$ GeV/c$^2$) but observe 550 events, due primarily
to $W$+jets production.  To enhance the relative contribution of events
from top quark decays, we employ two techniques. One method, denoted
$\ell$+jets/$\mu$, requires a  jet to be associated with a tag muon as
evidence of the semileptonic decay of a $b$ quark. A requirement on the
minimum separation between the muon and  the reconstructed jet  $\Delta
{\cal R}_{\rm jet} = \sqrt{\Delta \eta^2+\Delta \phi^2}$  defines this
association. The other method, denoted $\ell$+jets, is applied to events
without tag muons. It exploits the difference in event shape and
kinematics between $t\overline t$ and background. Selection criteria for
both methods are described in Table~\ref{tab:top}.  Note that the
requirements on  event shape variables are less stringent  for the
$\ell$+jets/$\mu$ analysis. 

\begin{figure}
\centerline{\psfig{figure=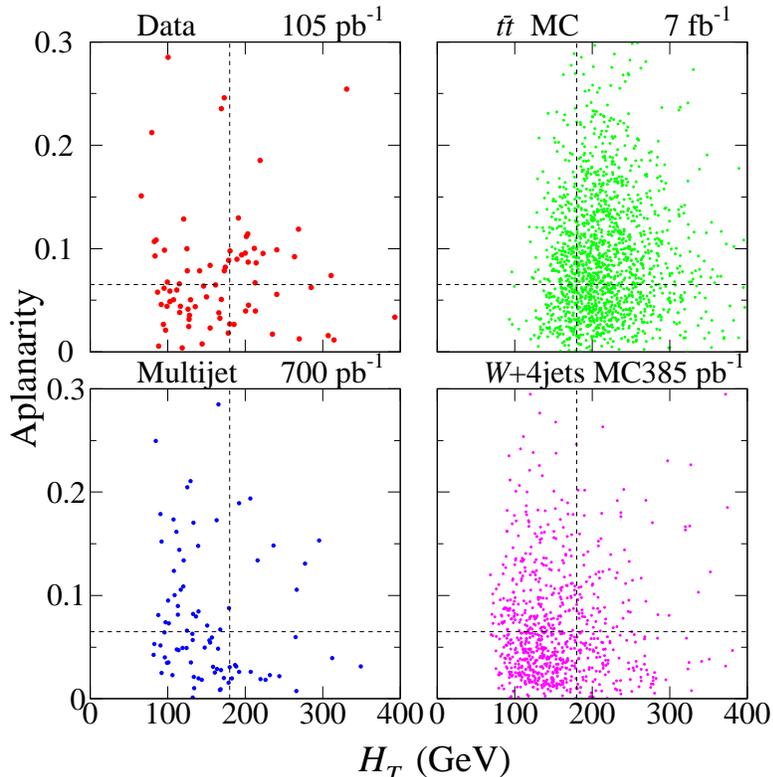,width=4.3in}}
\caption{Distributions of ${\cal A} $ vs. $H_T$
for $\ell$+jets data events  compared to
expectations  for  higher luminosity samples
of $t\overline t$ ($m_t=170$ GeV/c$^2$),
multijet, and $W$+4jets backgrounds.
The dashed lines represent the threshold
values used for the selection.}
\label{fig:ejaht}
\end{figure}

To select the optimal variables and their threshold values that  yield
the best precision for the measured cross section, we perform  an
optimization using a random grid search technique\cite{RGS}. We use a
Monte Carlo (MC) $t \overline t$ sample generated with $m_t=170$
GeV/c$^2$ to compute the expected signal event yield for various
cutoffs, while we determine the backgrounds using the methods described
below. Variables that provide significant discrimination between
$t\overline t$ events and backgrounds are $H_T$; the aplanarity ${\cal
A}$ computed  using $W$ boson and jet momenta in the laboratory
frame\cite{aplan}; and $E_T^L$, the scalar sum of the lepton $E_T$ and
{\hbox{$\rlap{\kern0.25em/}E_T$}}.  A requirement on the pseudorapidity
$\eta_W$ of the $W$ boson which decays leptonically\cite{topmassprl} is
imposed in the $\ell$+jets analysis to obtain better agreement between
background control samples from data and the $W$+jets MC samples. In
Fig.~\ref{fig:ejaht}, we show plots of the two kinematic variables
${\cal A} $ and $H_T$, after imposing all cuts except those on the
variables plotted, for our $\ell$+jets data sample, $t\overline t$ MC
and the two background sources: multijet and $W$+4 jets events. The cuts
indicated by the dashed lines provide a good separation between the
expected signal  and backgrounds. The optimized selection criteria 
listed in Table~\ref{tab:top} yield  9 $e$+jets, 10 $\mu$+jets, 5
$e$+jets/$\mu$, and 6 $\mu$+jets/$\mu$ events.

We gain increased acceptance for $t\overline t$ production through a
more inclusive channel, the $e \nu$ channel, which requires an isolated
high $E_T$ electron, {\hbox{$\rlap{\kern0.25em/}E_T$}}$>50$ GeV,
transverse mass of $e\nu$,  $M_T^{e\nu}>$115 GeV, and two or more jets
with $E_T>30$ GeV. The $e \nu$ channel contains top signal mainly from
dileptons and $e$+jets top decays which fail the standard kinematic
selection. Four events survive the $e \nu$ requirements listed in
Table~\ref{tab:top}.

\begin{table}[t]
\caption{Event yields}
\label{tab:topyields}
\begin{center}
\begin{tabular}{lccccc}
channel           & events          &  background   &  \multicolumn{3}{c}{expected signal}     \\
                  & observed        &               &  \multicolumn{3}{c}{$m_t$ (GeV/c$^2$)}  \\
                  &                 &               &  150       &    170     & 190        \\
\hline
dilepton          &  5              & 1.4$\pm$0.4   & 5.9$\pm$1.0 & 4.1$\pm$0.7 & 2.6$\pm$0.5 \\
$\ell$+jets       &  19             & 8.7$\pm$1.7   &18.3$\pm$6.3 &14.1$\pm$3.1 & 9.2$\pm$1.4 \\
$\ell$+jets/$\mu$ &  11             & 2.4$\pm$0.5   & 9.1$\pm$1.7 & 5.8$\pm$1.0 & 3.7$\pm$0.6 \\
$e\nu$            &  4              & 1.2$\pm$0.4   & 2.5$\pm$0.8 & 1.7$\pm$0.5 & 1.1$\pm$0.3 \\
\hline
total             &  39             &13.7$\pm$2.2   & 35.9$\pm$8.8 &25.7$\pm$4.6 &16.6$\pm$2.4 \\
\end{tabular}
\end{center}
\end{table}

For all channels, the number of $t \overline t$ events expected to pass
the selection criteria is calculated for  top quark masses between 140
and 200 GeV/c$^2$. Samples of $t \overline t$ decays to all possible
final states are produced with the {\sc herwig} event
generator\cite{HERWIG} and a {\sc geant} model of the D\O\
detector\cite{GEANT}. We filter MC events according to the same criteria
as used for data. Therefore, the acceptances include events with
$W\rightarrow\tau\nu$ decays that pass the selection cuts. The
acceptances computed from MC are refined  by incorporating lepton
selection efficiencies measured using $Z \to ee, \mu\mu$ data.
Table~\ref{tab:topyields}  lists the expected number of signal events,
computed using the $t \overline t$ production cross section of Ref.
\cite{Laenen}, for three top quark masses along with the number of
observed events. The errors quoted include  the uncertainty in the jet
energy scale, differences between the {\sc herwig} and {\sc
isajet}\cite{isajet} event generators, lepton identification, and
trigger efficiencies.

We distinguish between physics backgrounds, which have the same final
states as the signal process, and instrumental backgrounds in which
objects in the final state were misidentified. Instrumental backgrounds
for all channels are estimated entirely from data, using control samples
consisting of multijet events and the measured probability for
misidentifying a jet as a lepton\cite{topprd}. For the physics
backgrounds discussed below, the distributions for $W$+jets background
are modeled using the {\sc vecbos} event generator\cite{berends} which
is interfaced to {\sc herwig} to fragment the partons. The background
estimates for all analyses are summarized in Table \ref{tab:topyields}.

Sources for physics backgrounds depend  on the channel under
consideration. The main physics backgrounds to the dilepton channels are
$Z$ boson, Drell-Yan,  and vector boson pair  production. These  are
estimated by MC simulations, and corrected for  efficiencies measured in
collider data. In the $e\mu$ channel, the signal to background ratio is
$\approx$10:1, where about half of the total background is due to
$Z\to\tau\tau$ events. In the $\mu\mu$ channel, $Z$ decays are rejected
by a kinematic fit to the $Z\rightarrow\mu\mu$ hypothesis. The
$Z\rightarrow ee$ background is reduced by raising the cut on 
{\hbox{$\rlap{\kern0.25em/}E_T$}} to 40 GeV for dielectron masses within
12 GeV of  the $Z$ mass. The dominant physics background process for the
$e \nu$  channel is $W(\rightarrow e\nu)$+jets production and is
strongly suppressed by the large transverse mass requirement. To
estimate this background, we use the number of $W$+$\geq$ 2 jets events
observed in our data before the transverse mass cut and the rejection of
the $M_T^{e\nu}$ cut determined using $W$+2 jets MC. Contributions to
the uncertainty in the background include 12\% for variations in the jet
energy scale (15\% for $e\nu$), 10\% for uncertainties in the cross
sections used for MC samples, 15\% for modeling of $H_T$ and $H_T^e$
distributions in the MC, and typically 5\% for multiple interactions.
For the $\mu\mu$ channel there is an additional 10\% uncertainty for the
kinematic fit.

\begin{figure}
\centerline{\psfig{figure=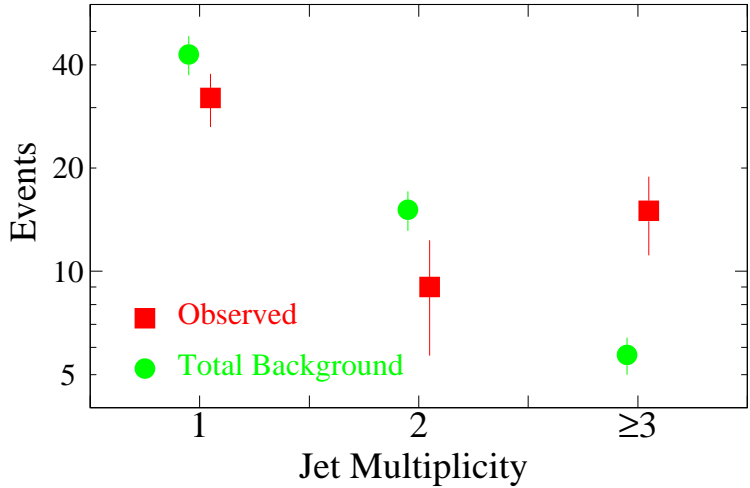,width=4.3in}}
\caption{Jet multiplicity spectrum of $\ell$+jets/$\mu$ events before
imposing event shape (${\cal A}, H_T$) criteria, compared to background estimates.}
\label{fig:njet_btag}
\end{figure}

In the $\ell$+jets channel, physics backgrounds arise mainly from
$W$+jets production. We estimate the $W$+jets background for events with
four or more jets by extrapolating from a $W$+jets data sample at low
jet multiplicities, assuming that the number of $W$+jets events falls
exponentially with the number of jets in the event ($N_{\rm jets}$
scaling)\cite{berends}. We have checked our $W$+jets data sample at jet
multiplicities between one and three, before event shape cuts (${\cal
A}, H_T$), and it supports this scaling law\cite{topprd}. We then apply
the survival probability for  event shape cuts which is determined to be
$9\pm 1\%$ from $W$+4 jets MC. The  uncertainty in the background
estimate includes a 10\% error on the validity of the $N_{\rm jets}$
scaling law (determined using $Z$+jets, $\gamma$+jets and multijet
control samples), 5\% for  jet energy scale variations, and 15\% for
differences in event shape variables between background and MC $W$+ 2
jets and $W$+ 3 jets samples.

The principal source of background in the $\ell$+jets/$\mu$ analysis is 
also $W$+jets production. We assume the heavy flavor content in $W$+jets
events is the same as in multijet events\cite{topprd}. The probability
of tagging a  jet in the absence of $t\overline t$ production is then
determined by the fraction of jets in multijet events that are tagged.
We parameterize the tagging rate as a function of jet $E_T$ and $\eta$.
By comparing the predicted and observed number of tags in several data
samples with jet $E_T$ thresholds varying from 20 to 85 GeV, we assign a
systematic uncertainty of 10\% to this procedure. We then apply this
tagging rate to each jet in a background dominated sample satisfying all
selection criteria in Table \ref{tab:top} except the $b$-tag
requirement. For the $\mu$+jets/$\mu$ final state, we reject
$Z(\rightarrow\mu\mu)$+jets events, where one of the muons is counted as
a tagging muon, by using a kinematic fit to the $Z$ decay hypothesis.
This residual background is estimated using a MC simulation. Figure
\ref{fig:njet_btag} shows the jet multiplicity spectrum of
$\ell$+jets/$\mu$ events and the background estimates before event shape
(${\cal A}, H_T$) cuts. There is good agreement for  1 and 2 jet
samples, while a clear excess  is observed at 3 or more jets, indicative
of $t\overline t$ production.

\begin{figure}
\centerline{\psfig{figure=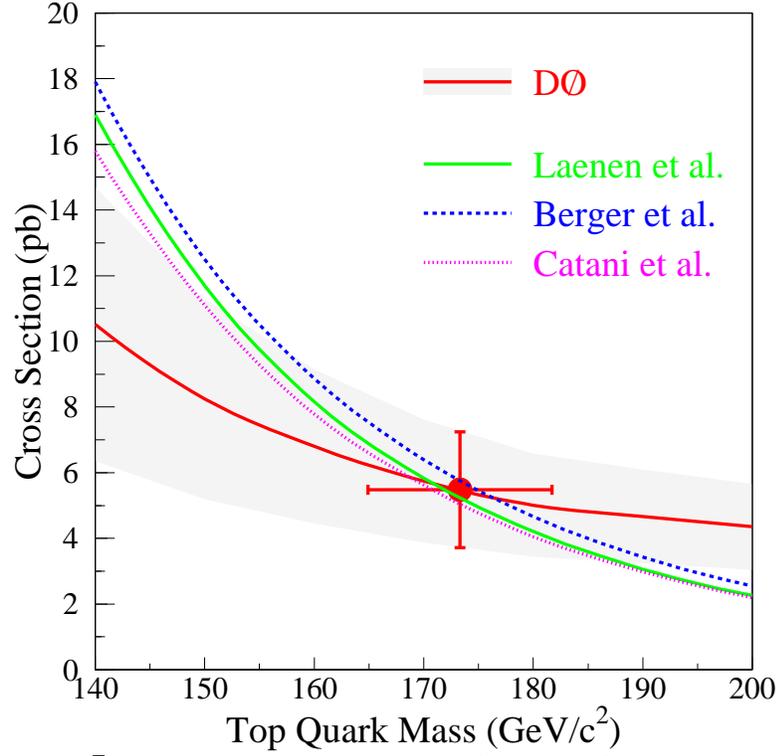,width=4.3in}}
\caption{Measured $t\overline t$ production cross section as a function
of $m_t$ (shaded band). The point with error bars
is the  cross section  for  the measured top quark mass  at D\O.
Three different theoretical estimates are also shown.}
\label{fig:xsect}
\end{figure}

Overall, 39 events satisfy the selection criteria. We expect
13.7$\pm$2.2 events from background sources and 24.2$\pm$4.1 $t\overline
t$ events, assuming $m_t=173$ GeV/c$^2$ and the predicted cross section
of Ref. \cite{Laenen}. The total acceptance for $t\overline t$ events
varies between 2.8\% and 4.9\% for top quark masses between 150 and 190 
GeV/c$^2$. Figure \ref{fig:xsect} shows the measured $t\overline t$
cross section versus top quark mass, compared to three theory
calculations\cite{Laenen,xsectheory1,xsectheory2}. The error band
accounts for statistical and systematic uncertainties, both in the
backgrounds and acceptances, and takes  account  of the correlations
among channels. The systematic uncertainty has a component due to $m_t$
dependent variations between MC generators (gen) used  to model top
production, while all other fractional systematic uncertainties are
$m_t$ independent.

We quote $\sigma_{t\overline t}$ at our  central value $m_t$ =
$173.3$~GeV/c$^2$\cite{topmassprl}. Cross section measurements for the
individual channels are consistent with each other; we measure
$6.3\pm3.3$ pb from dilepton and $e\nu$, $4.1\pm2.0$ pb from
$\ell$+jets, and $8.2\pm3.5$ pb from $\ell$+jets/$\mu$ events. Combining
them gives $\sigma_{t\overline t}$ = $5.5 \pm 1.4 ({\rm stat})\pm
0.9({\rm syst}) \pm 0.6 ({\rm gen})$ pb, in good agreement with the SM
predictions. Adding the three uncertainties in quadrature, we measure
the $t\overline t$ production cross section to be $5.5\pm1.8$ pb.

%
We thank the staffs at Fermilab and collaborating institutions for their
contributions to this work, and acknowledge support from the 
Department of Energy and National Science Foundation (U.S.A.),  
Commissariat  \` a L'Energie Atomique (France), 
State Committee for Science and Technology and Ministry for Atomic 
   Energy (Russia),
CNPq (Brazil),
Departments of Atomic Energy and Science and Education (India),
Colciencias (Colombia),
CONACyT (Mexico),
Ministry of Education and KOSEF (Korea),
CONICET and UBACyT (Argentina),
and the A.P. Sloan Foundation.

\end{document}